\documentclass[aps,prx,twocolumn,superscriptaddress,longbibliography]{revtex4-1}

\usepackage[english]{babel} 
\usepackage[latin1]{inputenc}
\usepackage[usenames,dvipsnames]{color}
\usepackage{graphicx}
\usepackage{amsmath,amssymb,amsfonts}
\usepackage[colorlinks=true,linkcolor=blue,urlcolor=blue,citecolor=blue]{hyperref}
\usepackage[percent]{overpic}

\begin{document}

\title{Levitons in helical liquids with Rashba spin-orbit coupling probed by a superconducting contact}

\author{Flavio Ronetti}
\affiliation{Dipartimento di Fisica, Universit\`a di Genova, Via Dodecaneso 33, 16146, Genova, Italy}
%
\author{Matteo Carrega}
\affiliation{NEST, Istituto Nanoscienze-CNR and Scuola Normale Superiore, Piazza S. Silvestro 12, I-56127 Pisa, Italy}

\author{Maura Sassetti}
\affiliation{Dipartimento di Fisica, Universit\`a di Genova, Via Dodecaneso 33, 16146, Genova, Italy}
\affiliation{CNR-SPIN, Via Dodecaneso 33, 16146, Genova, Italy}

\begin{abstract}
We consider transport properties of a single edge of a two-dimensional topological insulators, in presence of Rashba spin-orbit coupling, driven by two
external time-dependent voltages and connected to a thin superconductor. We focus on the case of a train of Lorentzian-shaped pulses, which are known to
generate coherent single-electron excitations in two-dimensional electron gas, and prove that they are minimal excitations for charge transport also in
helical edge states, even in the presence of spin-orbit interaction. Importantly, these properties of Lorentzian-shaped pulses can be tested computing
charge noise generated by the scattering of particles at the thin superconductor. This represents a novel setup where electron quantum optics experiments
with helical states can be implemented, with the superconducting contact as an effective beamsplitter. By elaborating on this configuration, we also
evaluate charge noise in a collisional Hong-Ou-Mandel configuration, showing that, due to the peculiar effects induced by
Rashba interaction, a non-vanishing dip at zero delay appears.
\end{abstract}

	\maketitle

\section{Introduction}
The exciting progress in quantum electronics gathered during the last decade has yielded an impressive level of experimental control such that even single-electron scale can be properly attained. In this context, a new research field, known as electron quantum optics (EQO)~\cite{Bocquillon14,Grenier11b}, has rapidly risen aiming at reproducing quantum optics experiment using single- or few-electron states propagating in solid state devices.

The ground-breaking achievements that paved the way for EQO have been the realizations of on-demand sources of coherent electrons~\cite{dubois13, grenier13, Misiorny18,Glattli16_pss,Bauerle18}. The first implementation of a single-electron source, known as mesoscopic capacitor, has been accomplished by periodically driving a quantum dot, thus alternatively emitting an electron and a hole along the ballistic channels of a quantum Hall system~\cite{buttiker93_JPCM,buttiker93_PLA,Feve07,Bocquillon12,Parmentier12,Ferraro15}. Another injection scheme, proposed by Levitov and co-workers ~\cite{levitov96,ivanov97,keeling06}, is based on the idea of applying to a quantum conductor a periodic train of quantized Lorentzian-shaped pulses, carrying an integer number of particles. Indeed, they predicted that this kind of voltage shape excites minimal single-electron excitations free from additional electron-hole pairs, then named Levitons~\cite{Ferraro18,Misiorny18,Moskalets16_PRL,Safi14,Dolcini17}. Compared to the mesoscopic capacitor, the emission of Levitons can be realized without the need for a precise fine-tuning of experimental parameters and is interestingly promising for miniaturization and scalability~\cite{Glattli16_pss,Glattli16_physE}. Due to their peculiar features, Lorentzian-shaped pulses have spurred a considerable number of theoretical proposals and experimental achievements, including the possibility to employ Levitons in quantum information processing~\cite{Dasenbrook15,Dasenbrook16_NJP,Dasenbrook16_PRL,Ferraro18b} or to reconstruct their single-electron wave-functions by means of a quantum tomography protocol~\cite{Grenier11,Ferraro13,Ferraro14,Jullien14}.

Among the most relevant EQO experiments, a prominent position is held by the Hanbury-Brown-Twiss (HBT) configuration, where a single source excites single-electron states along ballistic edge channels, which are partitioned against a beamsplitter, i.e. a quantum point contact (QPC), thus generating a shot noise signal employed to probe the single-electron nature of Levitons~\cite{Bocquillon12,Martin_Houches,Moskalets17,Rech16,Vannucci17,Ronetti18,Glattli16_physE,dubois13-nature,Acciai19b}. Another remarkable experimental configuration of EQO is provided by the Hong-Ou-Mandel (HOM) interferometer~\cite{Hong87}, where two sources of single electrons are placed on the opposite side of a scatterer, say the QPC and delayed by a tunable time~\cite{Jonckheere12,Ferraro18,Ferraro15,Bocquillon13,Glattli16_pss,dubois13-nature}. When two electrons collide at zero time delay at the QPC, charge noise is known to vanish at zero temperature, thus showing the so called Pauli dip, which is a consequence of the anti-bunching effect imposed by fermionic nature of electrons~\cite{Wahl14,Ferraro14,Freulon15,Marguerite16,Cabart18,Bellentani19,Ferraro14b,Glattli16_pss,dubois13-nature,Ronetti18b}. 

Propagation of coherent electron states can be properly attained in topologically protected edge channels of quantum Hall~\cite{Haldane17,Klitzing80,Girvin99,Haldane88} or quantum spin Hall systems~\cite{Bernevig06,Wu06,Dolcetto16rev,Calzona15b}. Topological protection guarantees ballistic transport regime, allowing for remarkable propagation length of single-electron states~\cite{Bendias18}. While there exist many proposals involving Levitons propagating in both integer and fractional quantum Hall systems, less attention has been paid to quantum spin Hall systems~\cite{Acciai19}, which, nevertheless, hold great promises for the peculiar spin properties. Quantum spin Hall effect arises along the edges of a two dimensional (2D) topological insulator (TI), where a pair of conductive edge states appear, with electrons with opposite spin polarization propagating in opposite directions, according to the so-called spin-momentum locking. These edge states are termed helical and their topological protection holds as long as time-reversal symmetry is preserved. The observation of a quantized conductance in HgTe/CdTe~\cite{Konig07,Brune12,Roth09} and InAs/GaSb~\cite{Knez11,Knez14,Du15} quantum wells provided the first experimental signatures for 2D TIs.

Singlet proximity superconductivity can be induced in helical edge states by connecting them to a conventional superconductor~\cite{Bours18,Bours19}, thus allowing for the transfer of electrons between them~\cite{Virtanen12,Adroguer10,Bocquillon17,Bocquillon17b,Guiducci19,Keidel19}. The superconducting proximity effect in helical edge channels has been already investigated in relation to several peculiar effects, such as the appearance of Majorana zero-energy states~\cite{Akhmerov09,Fu09,Rainis14,Park15,Fleckenstein18,Keidel18,Schulz19} and fractional Josephson effects. The low-energy physics of this proximity coupling can be described in terms of a tunnelling of Cooper pairs into the edge states, where they are split into electrons belonging to two channel propagating in opposite directions. This process effectively acts as a scattering mechanism for particles incoming at the superconductor, which can be thus employed as an analog of a beam splitter for Levitons. In this sense, a thin superconductor provides an alternative solution to the quantum point contact used so far in electron quantum optics experiments, such as HBT and HOM configurations.\\

The effect of Rashba spin-orbit coupling, which breaks the conservation of spin axis in edge state~\cite{Braunecker10,Strom10,Schmidt12,Crepin12,Klinovaja12,Klinovaja13,Geissler14,Kainaris14,Manchon15,Klinovaja15,Dolcini17,Mitrofanov17}, can introduce additional scattering processes even within right-moving and left-moving channels. The presence of Rashba interaction in 2D TIs could be ascribed to an inversion asymmetry mechanism in the heterostructure or to strain effect~\cite{Entin01,Vayrynen11,Ortix15,Gentile15,Ying16,Liu19}, that could be induced, for instance, by an external electric field~\cite{Qiao13,Wojcik14}.  In this way, the splitting of Cooper pairs can occur within the same channel, without any breaking of time-reversal symmetry. As a result, the presence of Rashba interaction makes the scattering dynamics at the superconductor even richer and more appealing~\cite{Virtanen12}.

In this paper, we shows that a novel setup with helical states and a superconducting element can be implemented for electron quantum optics experiments and that this would lead to peculiar results in presence of Rashba spin-orbit coupling. At this purpose, we focus on a single edge of a two-dimensional topological insulators, in presence of Rashba coupling, connected to a thin superconductor. Two external time-dependent voltages are applied to the system, thus allowing for the investigation of transport properties in this setup.
By solving the associated equation of motion, we provide the scattering matrix for particles at the superconductor, for any value of Rashba spin-orbit strength and proximity coupling constant. Then, we compute charge noise generated by the scattering of particles at the superconductor and, by focussing on the specific case of a single train of Lorentzian-shaped pulses, we prove that Levitons are minimal excitations for charge transport. Importantly, these properties of Lorentzian-shaped pulses still hold even in the presence of Rashba spin-orbit coupling.

Finally, we consider another paradigmatic case in EQO, investigating a collisional HOM configuration for Lorenzian-shaped pulses. Here, a non-vanishing dip at zero delay appears due to peculiar effects associated with Rashba interaction. Indeed, a signature that could assess the presence of Rashba spin-orbit interaction is important to determine the spin of emitted Levitons. When spin quantization axis is preserved, spin-momentum locking of edge state ensures that the spin of a right-moving Leviton is necessarily $\uparrow$. Nevertheless, in the presence of a finite Rashba coupling, spin is not a good quantum number and the probability of observing a right-moving Levitons with spin $\uparrow$ or $\downarrow$ is intimately related to the strength of Rashba spin-orbit strenght, which is therefore an important quantity to estimate. Another remarkable signature of Rashba interaction in this HOM configuration is that the charge noise for the case of a single Leviton is no longer universal with respect to temperature for finite values of Rashba spin-orbit strength, in contrast with previous literature on Hong-Ou-Mandel configurations with levitons. In passing, we notice that in on-going experiments on helical edge states of 2D TI in close proximity to superconductors~\cite{Bocquillon17,Bocquillon17b}, no clear evidence of strong $e-e$ interaction have been reported so far. This well justify our approach based on a non-interacting picture of helical edge states.\\

This paper is organized as follows. In Sec.~\ref{sec:model}, we present the setup under investigation. The associated equations of motion are solved in Sec.~\ref{sec:eq}, where we also derive the scattering matrix for the thin superconductor. Charge noise is evaluated in Sec.~\ref{sec:noise}. By focusing on the case of Levitons, we discuss our main findings in Sec.~\ref{sec:results}. Finally, we draw our conclusion in Sec.~\ref{sec:conlusions}. One Appendix contains technical details of our calculation.
\section{Model \label{sec:model}}
\begin{figure}
	\centering
	\includegraphics[width=\linewidth]{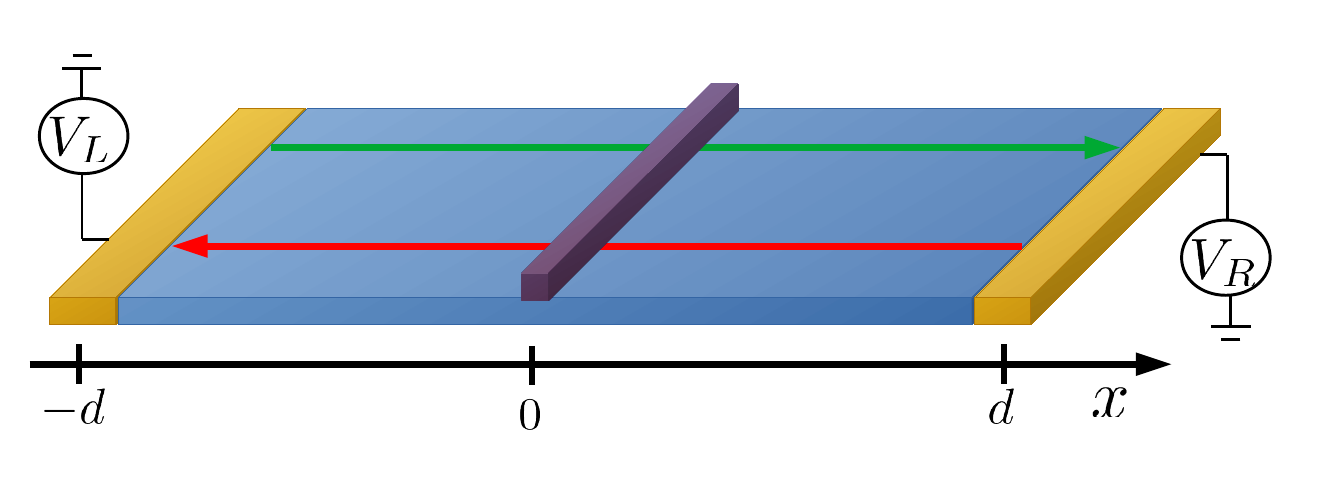}
	\caption{(Color online) Single edge of a 2D TI in a two-terminal geometry, in presence of a single thin superconductor (brown) placed above the system at $x=0$. A Kramer's pair of counterpropagating edge modes is formed by a right-moving (green) and left-moving (red) channels. In this scheme, edge states are drawn separately to easily distinguish them. The presence of a superconductor induces superconducting proximity effect in the edge states. Two time-dependent voltage drives, $V_L$ and $V_R$ are applied to left and right terminals, respectively, thus driving the system out of equilibrium.}
	\label{fig:setup}
\end{figure}

We consider a single edge of a 2D TI, where a Kramer pair of counterpropagating channels is formed.
Here, well-known spin-momentum locking property constraints the spin projection (up/down) and propagation direction (right/left) of edge modes. We take into account possible presence of Rashba spin-orbit coupling along the edge which can break spin-momentum locking. These Rashba components may be of intrinsic nature or can be induced by means of external electrostatic field, i.e. by acting on lateral gates~\cite{Scherubl16,Iorio19}. In addition, a single thin superconducting terminal is coupled to the edge states in the middle of the bar, see Fig.~\ref{fig:setup}.
Here, we assume that the thin superconductor is tunnel coupled to the 2D TI and induces singlet-type proximity correlations between electrons.
Since we are interested in studying transport properties, in presence of time-dependent external drives, we assume that the system is connected to two reservoirs, driven by external time-dependent gate potentials. The total system is described by the following Hamiltonian
\begin{equation}
H=H_{{\rm edge}}+H_{\Delta}+H_{{\rm g}},
\label{eq:ham_tot}
\end{equation}
where $H_{{\rm edge}}=H_{0}+H_{\rm so}$ describes non-interacting helical edge states in presence of Rashba interaction.
In terms of the spinor operator
\begin{equation}
\Psi(x)=\left(\begin{matrix}
\psi_{\uparrow}(x)\\\psi_{\downarrow}(x),
\end{matrix}
\right)
\label{eq:spinbasis}
\end{equation}
where $\psi_{\sigma}$ annihilates an electron with spin $\sigma=\uparrow / \downarrow$ in the position $x$, the free Hamiltonian of edge states with linear dispersion is given by (from now on we set $\hbar=1$)
\begin{equation}
H_{0}=-i v_F\int dx \Psi^{\dagger}(x)\sigma_z \partial_x\Psi(x),
\end{equation}
where $v_F$ is the Fermi velocity and $\sigma_z$ is the third Pauli matrix acting on the spin space.The uniform Rashba interaction is described by 
\begin{equation}
H_{{\rm so}}=-i\alpha \int dx  \Psi^{\dagger}(x)\sigma_y \partial_x\Psi(x),
\end{equation}
where $\alpha$ parametrizes Rashba spin-orbit strength and $\sigma_y$ is the second Pauli Matrix. For convenience, we assumed a uniform strength ($\alpha(x)=\alpha$), since it does not qualitatively affect any of our results. For later discussion, it is convenient to introduce the Rashba angle, defined as
\begin{equation}
\theta_{so} =\arctan\left(\frac{\alpha}{v_F}\right).
\end{equation}
Without Rashba spin-orbit interaction, the edge states reduce to the usual helical liquid characterized by spin-momentum locking, where electrons wiht spin $\uparrow$ ($\downarrow$) propagate in the right (left) direction. Due to the presence of Rashba coupling, the direction of motion of electrons is not locked to their spin projection and the spinor in Eq.~\eqref{eq:spinbasis} does not describe chiral modes.

Edge states are tunnel coupled to a superconducting terminal at position $x=0$, inducing proximity effect. In the limit where the superconducting gap $\Delta$ is the largest energy scale, an effective description of the superconducting term is given by the Hamiltonian~\cite{Virtanen12,Maslov96}
\begin{equation}
H_{\Delta}=\int dx \hspace{1mm} \overline{\Delta}(x) \psi_{\uparrow}(x)\psi_{\downarrow}(x)+\text{h.c.},
\end{equation}
where $\overline{\Delta}(x)$ describes the superconducting proximity gap induced in the edge states~\cite{Dolcetto12,vannucci15,ronetti16,Ronetti17}. In the following discussion, we will restrict to the case of a thin superconductor placed at position $x=0$, such that $\overline{\Delta}(x)=\overline{\Delta} \delta(x)$.
The constant  $\overline{\Delta}$ is the induced superconducting gap and its value is upper bounded by the bulk superconducting gap $\overline{\Delta}\leq \Delta$. The assumption that the superconducting gap is the largest energy scale prevent the tunneling of single quasiparticles between edge states and superconducting terminal. This low-energy effective Hamiltonian describes the injection of a Cooper pair into the edge states, that is successively separated in two electrons with opposite spin. Analogously, it takes into account the possibility that two electrons with opposite spin pairs up and tunnel into the superconducting terminal as a Cooper pair. \\

Finally, since we are interested in evaluating out-of-equilibrium properties of edge states, an external driving potential $V_{{\rm g}}(x,t)$ is capacitively coupled to the density of edge states $\rho(x)=\Psi^{\dagger}(x)\sigma_0 \Psi(x)$ (here $\sigma_0$ indicates the the identity matrix) via the Hamiltonian ($e>0$)~\cite{Dolcini16,Dolcini18}
\begin{equation}
H_{{\rm g}}=-e\int dx V_{{\rm g}}(x,t)\rho(x),
\end{equation}
where the space-time dependence of the external voltage is
\begin{equation}
\label{eq:gate_potential}
V_{{\rm g}}(x,t)=V_{L}(t)\theta\left(-x-d\right)+V_{R}(t)\theta\left(x-d\right),
\end{equation}
with $\pm d$ the position of right and left contacts, see Fig.~\ref{fig:setup}.
Time dependent voltages $V_{L/R}(t)=V_{L/R}(t+\mathcal{T})$ are assumed periodic with period $\mathcal{T}$ and corresponding frequency $\omega=\frac{2\pi}{\mathcal{T}}$.
Hereafter, we will keep general the form of $V_{L/R}(t)$ whenever possible, specifying it to the case of Lorentzian pulses while discussing the results.

\section{Equations of motion \label{sec:eq}}
Our goal is to compute charge current fluctuations generated in the presence of superconducting terminal and time-dependent external gate potentials.
Therefore, we have to find the expression of spinor field $\Psi(x,t)$ by solving the equations of motion derived from the total Hamiltonian in Eq.~\eqref{eq:ham_tot}. These are given by
\begin{align}
i\partial _{t}\psi_{\uparrow/\downarrow}&=\pm i v_F\partial_x\psi_{\uparrow/\downarrow}\pm  \alpha \partial_x\psi_{\downarrow/\uparrow}\nonumber\\&+e V_{{\rm g}}(x,t) \psi_{\uparrow/\downarrow}\pm \overline{\Delta} \delta(x)\psi^{\dagger}_{\downarrow/\uparrow}.\label{eq:eq_mot}
\end{align}
In order to solve the full equations of motion, it is useful to find before the chirally propagating modes in the regions away from $x=0$,  where it is placed the superconductor.
The effect of the thin superconductor can be taken into account by means of a \textit{scattering matrix} approach~\cite{Buttiker92,Buttiker93,Lesovik97,blanter00}, which provides the expression of outgoing fermionic modes in terms of incoming ones, properly matching wavefunctions.
\subsection{Solution without proximity effect}
We now focus on the region outside the superconductor one, i.e. $x\ne 0$.
It is useful to recast equations of motion in terms of a spinor expressed in the chiral basis. The latter quantities can be obtained from the spinor in Eq.~\eqref{eq:spinbasis} by applying a rotation parametrized by the Rashba angle $\theta_{so}$, defined as
\begin{equation}
\chi  (x)=\left(\begin{matrix}
\cos\left(\frac{\theta_{so}}{2}\right)&-i\sin\left(\frac{\theta_{so}}{2}\right)\\-i\sin\left(\frac{\theta_{so}}{2}\right)&\cos\left(\frac{\theta_{so}}{2}\right)
\end{matrix}\right) \Psi(x),
\label{eq:chiralbasis}
\end{equation} 
where $\chi(x)=\left(\begin{matrix}
\chi_{+}(x)\\ \chi_{-}(x)
\end{matrix}\right)$. In this chiral basis, the equations of motion in Eq.~\eqref{eq:eq_mot} becomes for $x\ne 0$
\begin{equation}
i\partial _{t}\chi_{\pm}=\pm i v\partial_x\chi_{\pm }+e V_{{\rm g}}(x,t) \chi_{\pm},
\end{equation}
with $v$ the renormalized velocity by Rashba spin-orbit interaction as $v=\sqrt{v_F^2+\alpha^2}$.\\
The solution of the above equations can be written as
\begin{equation}
\chi_{\pm}(x,t)=e^{-i \phi_{\pm}(x,t)}\gamma_{\pm}(x,t),
\label{eq:chiral_sol}
\end{equation}
where
\begin{equation}
\gamma_{\pm}(x,t)=\frac{1}{\sqrt{2\pi v}}\int dE e^{-i E(t\mp\frac{x}{v})} c^{(0)}_{\pm}(E)
\label{eq:chiral_sol0}
\end{equation}
are the solutions in absence of external gate potentials expressed using fermionic annihilation operators $c^{(0)}_{\pm}(E)$ and where we have introduced the phase factor
\begin{equation}
\label{eq:voltage_phase}
\phi_{\pm}(x,t)=e \int_{-\infty}^{t} dt' V_{{\rm g}}(x\mp v(t-t'),t').
\end{equation}
Fermionic operators $c^{(0)}_{\pm}(E)$ satisfy the following average values
\begin{align}
\left\langle{c^{(0)}_{\pm}}^{\dagger}(E')c^{(0)}_{\pm}(E)\right\rangle&=\delta(E-E')n(E),\\
\left\langle c^{(0)}_{\pm}(E'){c^{(0)}_{\pm}}^{\dagger}(E)\right\rangle&=\delta(E-E')(1-n(E)),
\end{align}
where $n(E)=(1+e^{\frac{E}{k_{{\rm B}} \theta}})^{-1}$ is the Fermi-Dirac equilibrium distribution at system temperature $\theta$  (assumed here to be the same for both reservoirs) and we set, for simplicity, the chemical potential to zero.

By focusing on the regions of propagation between contacts and superconductor, namely $-d<x<0$ for right-moving electrons ($\chi_{+}$) and $0<x<d$ for left moving electrons ($\chi_{-}$), and considering the space-dependence of $V_{{\rm g}}$ (see Eq.~\eqref{eq:gate_potential}), the phase factor in Eq.~\eqref{eq:voltage_phase} becomes
\begin{equation}
\phi_{\pm}(x,t)=e \int_{-\infty}^{t\mp \frac{x\mp d}{v}} dt' V_{L/R}(t').
\label{eq:phi2}
\end{equation}
Notice that, even though the gate potential is coupled to the total density operator $\rho$, only electrons chirally propagating to the right (left) are affected by $V_L$ ($V_R$).\\
By using Eq.~\eqref{eq:phi2}, the exponential in Eq.~\eqref{eq:chiral_sol} can be conveniently rewritten as
\begin{equation}
e^{-i \phi_{\pm}(x,t)}=\sum_{l}p_{l,L/R}e^{-i (l+q_{L/R})\omega\left(t\mp \frac{x\mp d}{v}\right)},
\label{eq:phi3}
\end{equation}
where we have introduced the Fourier coefficients associated to voltages $V_{L}$ and $V_R$
\begin{equation}
p_{l,L/R}=\int_{0}^{\mathcal{T}}\frac{dt}{\mathcal{T}}e^{-i e \int_{0}^{t} dt' V_{L/R}(t')}e^{i q_{L/R}\omega t}e^{il \omega t},
\label{eq:pl1}
\end{equation}
and we have defined
\begin{equation}
q_{L/R}=e\int_{0}^{\mathcal{T}}\frac{dt}{2\pi}V_{L/R}(t),
\end{equation}
which represent the number of particles emitted by left and right voltage per period into $+$ and $-$ chiral channel, respectively. Coefficients in Eq.~\eqref{eq:pl1} represent the probability amplitude for electrons to absorb ($l>0$) or emit ($l<0$) a photon at energy $\omega$ and, for this reason, are called \textit{photo-assisted coefficient}~\cite{dubois13,Vannucci18,Ronetti19}.\\
By defining the following fermionic annihilation operator
\begin{equation}
c_{\pm}(E)=\sum_{l}p_{l,L/R}e^{i l \omega \frac{d}{v}}c^{(0)}_{\pm}\left(E+\omega\left(l+q_{L/R}\right)\right),
\end{equation}
the right- and left-moving chiral modes, corresponding to the modes incoming to the superconductor, respectively, from $x<0$ and $x>0$, are given by
\begin{equation}
\chi^{(i)}_{\pm}(x,t)\equiv \chi_{\pm}(x,t)=\frac{1}{\sqrt{2\pi v}}\int dE e^{-i E(t\mp\frac{x}{v})} c_{\pm}(E).
\label{eq:chiral_sol2}
\end{equation}
In passing, it is useful to provide the expressions for average values involving fermionic operators $c_{\pm}(E)$ in terms of photo-assisted coefficients $p_{l,L/R}$
\begin{align}
&\left\langle c^{\dagger}_{\pm}(E')c_{\pm}(E)\right\rangle=\sum_{m,l}p_{m+l,L/R}^{*}p_{l,L/R}e^{-im \omega \frac{d}{v}}\nonumber\times\\&\delta\left(E'-E+\omega m\right)n\left(E+\omega(l+q_{L/R})\right),\label{eq:av_c1}\\
&\left\langle c_{\pm}(E')c^{\dagger}_{\pm}(E)\right\rangle=\sum_{m,l}p_{m+l,L/R}p^{*}_{l,L/R}e^{im \omega \frac{d}{v}}\nonumber\times\\&\delta\left(E'-E+\omega m\right)n\left(-E-\omega(l+q_{L/R})\right).\label{eq:av_c2}
\end{align}
\subsection{Scattering matrix at the superconductor}
Having found the expressions for incoming chiral modes, we can consider the presence of the thin superconducting terminal at $x=0$ and find a solution to the full equations of motion in Eq.~\eqref{eq:eq_mot}. In order to tackle this problem, chiral modes outgoing from the superconductor can be connected to the incoming ones (given in Eq.~\eqref{eq:chiral_sol2}) by means of a scattering matrix. Outgoing fermionic modes can be formally defined as
\begin{equation}
\chi^{(o)}_{\pm}(x,t)\equiv\frac{1}{\sqrt{2\pi v}}\int dE e^{-i E(t\mp\frac{x}{v})} d_{\pm}(E),
\label{eq:chiral_out}
\end{equation}
where $d_{\pm}(E)$ annihilate an outgoing electronic mode at energy $E$. To obtain the scattering matrix for this problem, it is useful to integrate Eq.~\eqref{eq:eq_mot} around $x=0$. One has
\begin{align}
i v_F \left(\psi_{\uparrow/\downarrow}(0^+)-\psi_{\uparrow/\downarrow}(0^-)\right)&+ \alpha \left(\psi_{\downarrow/\uparrow}(0^+)-\psi_{\downarrow/\uparrow}(0^-)\right)=\nonumber\\& = \overline{\Delta} \frac{\psi^{\dagger}_{\downarrow/\uparrow}(0^+)+\psi^{\dagger}_{\downarrow/\uparrow}(0^-)}{2},
\label{eq:eq_mot2}
\end{align}
where $V_{{\rm g}}(x,t)$ does not appear since it is zero around the origin. In order to recast the above equations of motion in terms of chiral modes, let us observe that, by inverting Eq.~\eqref{eq:chiralbasis} and using the definition of incoming and outgoing chiral modes, one has
\begin{align}
\psi_{\uparrow}(0^{\pm},t)&=\cos\left(\frac{\theta_{so}}{2}\right)\chi^{(o/i)}_{+}(0,t)+i\sin\left(\frac{\theta_{so}}{2}\right)\chi^{(i/o)}_{-}(0,t),\\\psi_{\downarrow}(0^{\pm},t)&=i\sin\left(\frac{\theta_{so}}{2}\right)\chi^{(o/i)}_{+}(0,t)+\cos\left(\frac{\theta_{so}}{2}\right)\chi^{(i/o)}_{-}(0,t).
\end{align}
Note that, due to the presence of both creation and annihilation operators, one has to consider also the complex conjugate equations of Eq.~\eqref{eq:eq_mot2} in order to derive the complete scattering matrix. By substituting these expression in terms of chiral modes back into Eq.~\eqref{eq:eq_mot2} and integrating over $\int dt e^{i \epsilon t}$, we can find a linear system of equations in the energy space connecting incoming fermion operators to outgoing ones as
\begin{equation}
\left(\begin{matrix}
d_{+}(\epsilon)\\d_{-}(\epsilon)\\d^{\dagger}_{+}(-\epsilon)\\d^{\dagger}_{-}(-\epsilon)
\end{matrix}\right)=\mathcal{M}\left(\begin{matrix}
c_{+}(\epsilon)\\c_{-}(\epsilon)\\c^{\dagger}_{+}(-\epsilon)\\c^{\dagger}_{-}(-\epsilon)
\end{matrix}\right),
\label{eq:scattering}
\end{equation}
with $\mathcal{M}$ the $4\times 4$ scattering matrix given by
\begin{equation}
\mathcal{M}=\left(\begin{matrix}
	t_{ee} & 0 & t_{eh} & r_{eh}\\ 0 & t_{ee} & r_{eh} & t_{eh}\\t_{eh} &r^{*}_{eh} & t_{ee} & 0\\r^{*}_{eh} & t_{eh} & 0 & t_{ee} \\
	\end{matrix}\right),\label{eq:scattering_m1}
	\end{equation}
	where
	\begin{widetext}
	\begin{align}
	t_{ee}&=\frac{v^2-\left(2\cos\theta_{so}-1\right)\alpha^2+\left(\frac{\overline{\Delta}}{2}\right)^2}{\sqrt{\left[v^2-\left(2\cos\theta_{so}-1\right)\alpha^2+  \left(\frac{\overline{\Delta}}{2}\right)^2\right]^2 + v^2\overline{\Delta}^2 \left(1-\alpha\sin\theta_{so}\right)^2 + v^2\overline{\Delta}^2 \sin\theta_{so}^2}},\label{eq:tr_ee}\\
	r_{eh}&=\frac{i v\overline{\Delta} \left(1-\alpha\sin\theta_{so}\right)}{\sqrt{\left[v^2-\left(2\cos\theta_{so}-1\right)\alpha^2+  \left(\frac{\overline{\Delta}}{2}\right)^2\right]^2 + v^2\overline{\Delta}^2 \left(1-\alpha\sin\theta_{so}\right)^2 + v^2\overline{\Delta}^2 \sin\theta_{so}^2}},\label{eq:r_eh}\\
	t_{eh}&=\frac{v\overline{\Delta} \sin\theta_{so}}{\sqrt{\left[v^2-\left(2\cos\theta_{so}-1\right)\alpha^2+  \left(\frac{\overline{\Delta}}{2}\right)^2\right]^2 + v^2\overline{\Delta}^2 \left(1-\alpha\sin\theta_{so}\right)^2 + v^2\overline{\Delta}^2 \sin\theta_{so}^2}}.\label{eq:tr_eh}
	\end{align}
	\end{widetext}
	These coefficients represent the probability amplitude for each different scattering process that can occur at the superconductor. There are three possible events that can occur at the superconductor when a particle is incoming. In the first case, corresponding to probability amplitude $t_{ee}$, the electron (the hole) is unaffected by the presence of the superconductor and is transmitted without changing the sign of its charge. In the remaining scattering processes, which are taken into account by $r_{eh}$ and $t_{eh}$, the incoming electron (hole) effectively comes out of the superconductor as a hole. Indeed, these processes are a consequence of the singlet-type proximity correlations induced along the edge states~\cite{Virtanen12}. For the reflection process ($r_{eh}$), the incoming particle pairs up with an identical particle from the other channel and the so-formed Cooper pair is transferred to the superconductor. This process can be effectively described as the reflection of an electron into a hole, or viceversa. In the case of transmission ($t_{eh}$), the Cooper pair is formed by two identical particles belonging to the same channel, thus giving rise to an effective transmission amplitude for the process that converts electrons into holes and its opposite counterpart. The latter process is allowed solely in the presence of Rashba interaction, that, by breaking the conservation of spin quantization axis, can induce singlet-type correlation even between electrons flowing in the same channel. Indeed, when Rashba spin-orbit strength is zero ($\alpha=0$), the probability amplitude for transmission of an electron (a hole) as a hole (an electron) vanishes, since $\left(t_{eh}\right)_{\theta_{so}=0}=0$, and only the two remaining amplitudes are finite
	\begin{align}
	\left(t_{ee}\right)_{\alpha=0}&=\frac{v^2+\left(\frac{\overline{\Delta}}{2}\right)^2}{\sqrt{\left[v^2+\left(\frac{\overline{\Delta}}{2}\right)^2\right]^2 + v^2 \overline{\Delta}^2}},\label{eq:tr_ee0}\\
	\left(r_{eh}\right)_{\alpha=0}&=\frac{i v\overline{\Delta}}{\sqrt{\left[v^2+\left(\frac{\overline{\Delta}}{2}\right)^2\right]^2 + v^2 \overline{\Delta}^2}}.\label{eq:r_eh0}
	\end{align}
	Moreover, it is also useful to discuss the limit of vanishing superconducting proximity effect. In this case, i.e for $\overline{\Delta}=0$, one finds that $t_{eh}=r_{eh}=0$ and $t_{ee}=1$, thus implying that $\mathcal{M}=\mathbf{I}_4$. As a consequence, Eq.~\eqref{eq:scattering}  becomes a simple identity between $d_{\pm}(\epsilon)$ and $c_{\pm}(\epsilon)$, showing that the presence of the superconductor is crucial to obtain a finite scattering between different chiral modes.\\ 
	Finally, we remark that $\mathcal{M}$ is completely independent of energy. This is a consequence of the assumption of a very thin superconductor that translates into the condition of a scattering between modes localized in a single point, i.e. $x=0$. 
\section{Calculation of noise \label{sec:noise}}
As a consequence of the scattering mechanisms described in the previous Section, fluctuations of charge current are generated~\cite{Martin_Houches,blanter00,Ferraro14c}. In this Section we derive the expression for these fluctuations at any order in Rashba strength $\alpha$ and proximity effect coupling constant $\overline{\Delta}$. We still keep general the form of $V_{L/R}(t)$, which will be specified in the next section. Charge current operator entering reservoir $R/L$ ($x=\pm d$), in terms of chiral edge modes, are defined as 
\begin{align}
&J_{R/L}(t)=\nonumber\\&=e v \left({\chi^{(o)}_{\pm }}^{\dagger}(\pm d,t)\chi^{(o)}_{\pm}(\pm d,t)-{\chi^{(i)}_{\mp }}^{\dagger}(\pm d,t)\chi^{(i)}_{\mp }(\pm d,t)\right).\label{eq:curr}
\end{align}
The zero-frequency charge noise is given by
\begin{equation}
\mathcal{S}_{\alpha \beta}=\int_{0}^{\mathcal{T}}\frac{dt}{\mathcal{T}}\int dt'\left[\left\langle J_{\alpha}(t')J_{\beta}(t)\right\rangle-\left\langle J_{\alpha}(t')\right\rangle \left\langle J_{\beta}(t)\right\rangle\right],
\end{equation}
where $\alpha/\beta=R,L$ are labels for the two reservois. Since, according to the unitarity of scattering matrix, it is possible to prove that
\begin{equation}
\label{eq:noise_rel}
\mathcal{S}_{RL}=\mathcal{S}_{LR}=-\mathcal{S}_{RR}=-\mathcal{S}_{LL},
\end{equation}
we will focus only on the cross-correlator $\mathcal{S}_{RL}$ and define the charge noise as $\mathcal{S}_{C}\equiv\mathcal{S}_{RL}$.\\
Performing standard calculations, whose details are given in Appendix \ref{app:noise}, we evaluate explicitly $\mathcal{S}_{C}$. Obtaining
\begin{equation}
\mathcal{S}_C=A_{\theta}(\alpha,\overline{\Delta})\mathcal{S}_{\theta}+A_1(\alpha,\overline{\Delta})\mathcal{S}_1+A_2(\alpha,\overline{\Delta})\mathcal{S}_2,\label{eq:noise_fin}
\end{equation}
where $A_{\theta}$, $A_1$ and $A_2$ are dimensionless coefficients whose expression in terms of elements of the scattering matrix are
\begin{align}
A_{\theta}&=\left|r_{eh}t_{eh}\right|^2+\left|r_{eh}t_{ee}\right|^2-2(\left|t_{eh}\right|^2+\left|t_{ee}\right|^2),\\
A_1&=\left|t_{ee}\right|^2\left|r_{eh}\right|^2,\label{eq:A1}\\
A_2&=\left|t_{eh}\right|^2\left|r_{eh}\right|^2\label{eq:A2},
\end{align}
while the three contributions to noise are given by
\begin{align}
\mathcal{S}_{\theta}&=\mathcal{S}_0\sum_{s=\pm}\int \frac{dE'}{2\pi} \int \frac{dE}{2\pi} \left\langle c_{s}(E')c^{\dagger}_s(E)\right \rangle \left\langle c^{\dagger}_{s}(E')c_s(E)\right \rangle,\label{eq:thermal}\\
\mathcal{S}_1&=\mathcal{S}_0\int \frac{dE'}{2\pi} \int \frac{dE}{2\pi} \Bigg(\left\langle c_{+}(E')c^{\dagger}_+(E)\right \rangle \left\langle c_{-}(-E')c^{\dagger}_{-}(-E)\right \rangle+\nonumber\\&+\left\langle c^{\dagger}_{+}(E')c_+(E)\right \rangle \left\langle c^{\dagger}_{-}(-E')c_{-}(-E)\right \rangle\Bigg),\\
\mathcal{S}_{2}&=\mathcal{S}_0\sum_{s=\pm}\int \frac{dE'}{2\pi} \int \frac{dE}{2\pi}\left\langle c_{s}(E')c^{\dagger}_s(E)\right \rangle \left\langle c^{\dagger}_{-s}(E')c_{-s}(E)\right \rangle,
\end{align}
where the quantity $\mathcal{S}_0=\frac{e^2}{\mathcal{T}}$ has been defined. Let us observe that all the effects of Rashba spin-orbit coupling and superconductivity are encoded in the three coefficient $A_{\theta}$, $A_1$ and $A_2$, while the three components of charge noise are sensitive only to temperature and external voltages.\\
Some comments about the physical origin of these three contributions to charge noise are in order. The first one arises due to correlations between particles emitted from the same reservoir. Its expression in Eq.~\eqref{eq:thermal} can be further simplified by using Eqs. \eqref{eq:av_c1} and \eqref{eq:av_c2}. We arrive at
\begin{equation}
\label{eq:thermal_noise}
\mathcal{S}_{\theta}=2\mathcal{S}_0\int \frac{dE}{2\pi} n(E)\left(1-n(E)\right)=2\mathcal{S}_0 k_{\rm B} \theta,
\end{equation}
Therefore, one concludes that $\mathcal{S}_{\theta}$ corresponds to the thermal noise. This quantity is completely unrelated to the properties of time-dependent applied voltages and therefore will be neglected in the following discussion.\\
The other two contributions stem from correlations induced by particle exchange among the two reservoirs in presence of external drives. The different scattering processes that originate the fluctuations associated with $\mathcal{S}_1$ and $\mathcal{S}_2$ can be also deduced from the scattering matrix elements appearing in their corresponding coefficients $A_1$ and $A_2$. Both coefficients resemble the usual partition probability in terms of reflection and transmission coefficients. Nevertheless, the noise contribution $\mathcal{S}_1$ is generated by partitioning of electrons (holes) which can be probabilistically transmitted as themselves of reflected with as holes (electrons). On the other hand, the remaining noise contribution $\mathcal{S}_2$ is associated with a partitioning of electrons (holes) which are always converted by the superconductor into holes (electrons) either during a transmission or a reflection. These contributions can be expressed in terms of Fermi distribution function as 
\begin{align}
\mathcal{S}_1&=2\mathcal{S}_0\sum_{k}\left|p_{k,L+R}\right|^2\nonumber\\&\times\int \frac{dE}{2\pi \omega}\hspace{2mm} n\left(E+\left(k+q_L+q_R\right)\omega\right)n(-E),\\
\mathcal{S}_2&=2\mathcal{S}_0\sum_{k}\left|p_{k,L-R}\right|^2\nonumber\\&\times\int \frac{dE}{2\pi\omega}\hspace{2mm} n\left(E+\left(k+q_L-q_R\right)\omega\right)n(-E),
\end{align}
where we have introduced the photo-assisted coefficient $p_{k,L\pm R}$ associated to the effective drive corresponding to the sum and the difference of $V_L$ and $V_R$
\begin{equation}
p_{k,L\pm R}=\int_{0}^{\mathcal{T}}\frac{dt}{\mathcal{T}}e^{i k \omega t}e^{-ie \int_{0}^{t} dt' \left(V_L(t')\pm V_R(t')\right)}e^{i \left(q_L\pm q_R\right)\omega t}.
\end{equation}
In this form, it is manifest that $\mathcal{S}_1$ depends solely on the sum between the two external voltages, while $\mathcal{S}_2$ is governed by their difference. This property is a consequence of the physical origin previously described for these two noise terms.\\
Integrating over energies we obtain
\begin{align}
\mathcal{S}_1&=2\mathcal{S}_0 k_{\rm B} \theta\sum_{k}\left|p_{k,L+R}\right|^2\mathcal{F}\left(\left(k+q_L+q_R\right)\frac{\omega}{k_{\rm B} \theta}\right),\\
\mathcal{S}_2&=2\mathcal{S}_0 k_{\rm B} \theta \sum_{k}\left|p_{k,L-R}\right|^2\mathcal{F}\left(\left(k+q_L-q_R\right)\frac{\omega}{k_{\rm B} \theta}\right),
\end{align}
where, for notational convenience, we have defined the function $\mathcal{F}(x)=x \coth(x)$.\\ 
It is instructive to provide the limits of coefficients $A_1$ and $A_2$ (see Eqs. \eqref{eq:A1} and \eqref{eq:A2}) for vanishing superconducting coupling parameter $\overline{\Delta}$ and Rashba interaction strength $\alpha$. In the first case, we know that $\mathcal{M}=\mathbf{I}_4$ and, as a consequence, all off-diagonal elements are zero. For this reason,
\begin{align}
A_1(\alpha,\overline{\Delta}=0)=A_2(\alpha,\overline{\Delta}=0)=0,
\end{align}
meaning that superconductivity is necessary to observe a finite charge noise in addition to thermal noise $\mathcal{S}_{\theta}$. In the other limiting case, one has that $\left(t_{eh}\right)_{\alpha=0}=0$, thus
\begin{align}
&A_1(\alpha = 0 , \overline{\Delta}) = \frac{v^2 \overline{\Delta}^2\left|v^2+\left(\frac{\overline{\Delta}}{2}\right)^2\right|^2 }{\left[v^2+\left(\frac{\overline{\Delta}}{2}\right)^2\right]^2 + v^2 \overline{\Delta}^2}\\ &A_2(\alpha=0,\overline{\Delta})=0,
\end{align}
which proves that the \textit{simultaneous} presence of two terms in charge noise, generated by different scattering processes, is entirely due to the presence of Rashba spin-orbit coupling.\\
\begin{figure}
	\centering
	\includegraphics[width=\linewidth]{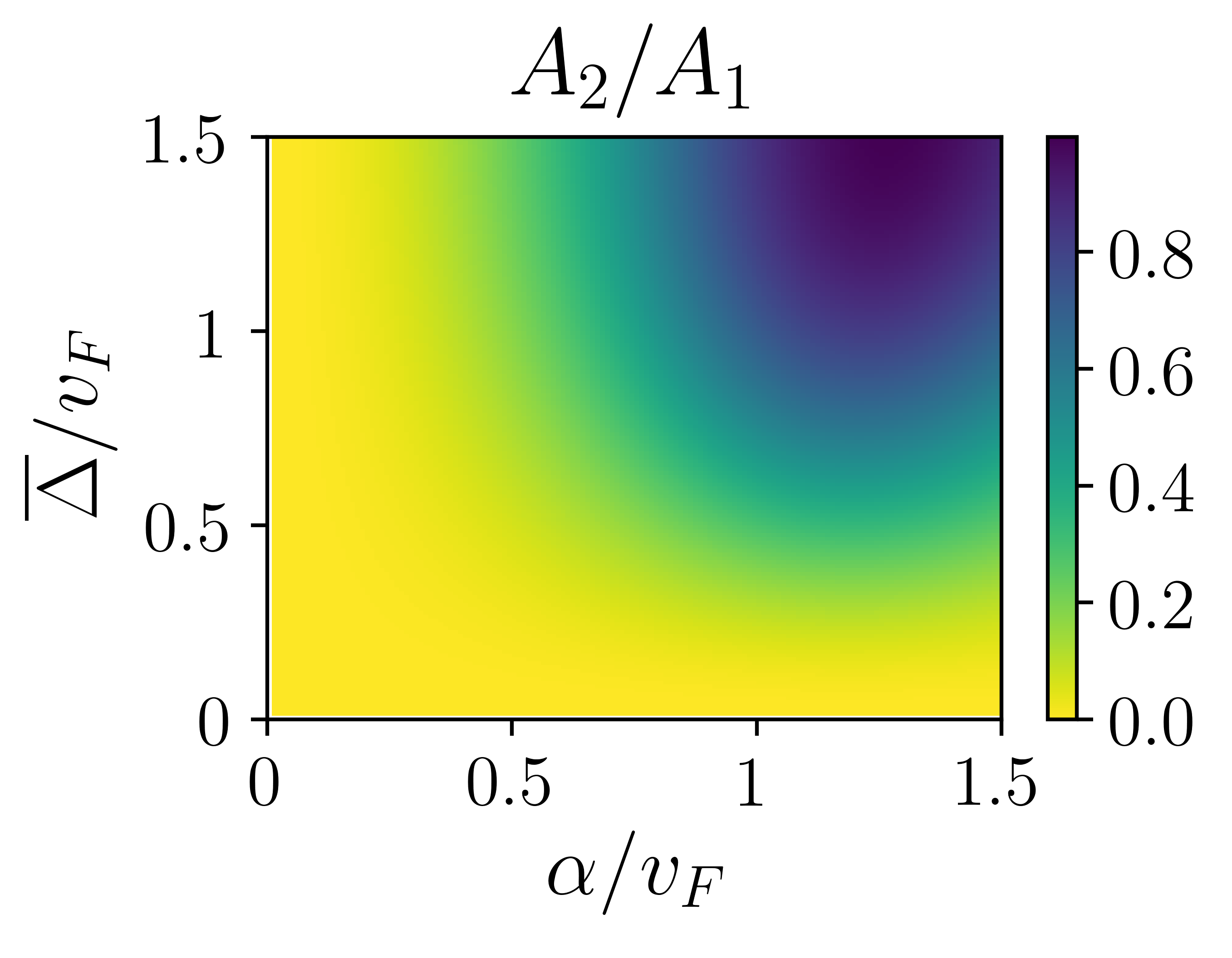}
	\caption{(Color online) Ratio $\frac{A_2}{A_1}$ as a function of Rashba spin-orbit strength $\alpha$ and proximity induced superconducting gap $\overline{\Delta}$. This ratio shows the relative weight between the two coefficients appearing in the charge noise $A_2$, which is finite only in presence of Rashba spin-orbit interaction, and $A_1$.}
	\label{fig:coefficient}
\end{figure}In Fig.~\ref{fig:coefficient}, the ratio $\frac{A_2}{A_1}$ as a function of $\alpha$ and $\overline{\Delta}$ is plotted, in order to understand the behaviour of these coefficients away from the above described limits. In general, as Rashba spin-orbit strength is increased this ratio is enhanced, meaning that the role of noise contribution $S_2$ associated with $A_2$ becomes more significant. Moreover, this enhancement is stronger as the proximity induced gap becomes larger.\\
\section{Results and discussion \label{sec:results}}
In this Section, we exploit our calculations to provide evidence for the emission of single-coherent electrons along helical edge states, even in the presence of Rashba interaction, by using a periodic train of Lorentzian-shaped pulses. The expression for the latter signal is given by
\begin{equation}
V_{\rm lor}(t)=\frac{V_0}{\pi}\sum_{k=-\infty}^{+\infty}\frac{\eta}{\eta^2+\left(\frac{t}{\mathcal{T}}-k\right)^2}, \label{eq:lor}
\end{equation}
where $V_0$ is a voltage amplitude and $\eta$ is the dimensionless half width at half maximum of each single Lorentzian pulse. Whenever each pulse carry an integer number of charge during each period $\mathcal{T}$, this specific form of external drive generates, along the edge states of quantum conductors, single-electron excitations, called Levitons, which can travel free from additional electron-hole pairs. This property has been shown theoretically for quantum Hall systems and experimentally for a non-interacting two-dimensional electron gas. The setup usually employed to prove that Levitons are minimal excitation states is the so called Hanbury-Brown-Twiss (HBT) configuration, where a single drive is applied to a quantum conductor in a quantum point contact (QPC) geometry~\cite{dubois13-nature,Rech16}. Here, we want to demonstrate that Levitons are minimal excitations even for helical edge states in the presence of Rashba spin-orbit coupling and show that evidence for this result could be achieved in a HBT-like configuration, i.e. with a single drive applied to the system, where instead of the partitioning at the QPC one can employ the scattering induced by a thin superconductor tunnel-coupled to the helical edge. To further prove the peculiarity of Lorentzian-shaped pulses, we compare it with a cosine drive
\begin{align}
V_{\rm cos}(t)&=V_0\left(1+\cos\left(2\pi \frac{t}{\mathcal{T}}\right)\right),
\end{align}
being a representative for all non-optimal drives. The form for photo-assisted coefficient of these two voltages are given, for instance, in Ref. \cite{dubois13}.
\subsection{Minimal excitation states} 
The HBT setup could be implemented in our system by considering the situation where $V_R(t)=0$. In this case, at zero temperature, charge noise becomes
\begin{equation}
\mathcal{S}^{HBT}_C\equiv \mathcal{S}_C\left[V_L(t),0\right]=2 \mathcal{S}_0\left(A_1+A_2\right)\sum_{k}\left|p_{l,L}\right|^2\left|k+q_L\right|,
\end{equation}
where the squared brackets indicate the chosen voltage configuration. Let us observe that the two partitioning processes associated with coefficients $A_1$ and $A_2$ contributes equally to charge noise in this configuration.\\
In order to assess that Levitons are minimal excitations for charge transport, one has to show that the number of additional electron-hole pairs generated by a Lorentzian drive vanishes for integer values of $q_L$. At zero temperature, the total number of excitations generated by a generic voltage drive $V_L(t)$ is given by $N_{exc}=N_e+N_h$, where
\begin{align}
N_e=\int_{-\infty}^{0}d\epsilon\left\langle c_+^{\dagger}(\epsilon)c_+(\epsilon)\right\rangle=\sum_{l\ge-q_L}\left|p_{l,L}\right|^2\left|l+q_L\right|,\label{e:Ne}\\
N_h=\int_{0}^{+\infty}d\epsilon\left\langle c_+(\epsilon)c_+^{\dagger}(\epsilon)\right\rangle=\sum_{l\le-q_L}\left|p_{l,L}\right|^2\left|l+q_L\right|,\label{e:Nh}
\end{align}
where $N_e$ and $N_h$ are, respectively, the total number of electrons and holes generated. Since the number of single electrons emitted by the voltage $V_L(t)$ is equal to $q_L$, the amount of unwanted electron-hole pairs is given by
\begin{equation}
\label{eq:exc}
\Delta N_{exc}=N_{exc}-q_L=\sum_{l}\left|p_{l,L}\right|^2\left|l+q_L\right|-q_L.
\end{equation}
In analogy with previous works on Levitons, we can show that the above quantity is linked to a transport property, called excess noise, which is defined as~\cite{Glattli16,Glattli16_pss,dubois13-nature,glattli2016method}
\begin{equation}
\Delta \mathcal{S}_C\equiv \mathcal{S}^{HBT}_C-\mathcal{S}^{DC}_C,
\end{equation}
where $\mathcal{S}^{DC}_C=\mathcal{S}_C\left[\frac{e q_L}{\omega},0\right]$ is the charge noise generated by a purely DC drive given by the DC part of $V_L$ and whose expression is
\begin{equation}
\mathcal{S}^{DC}_C=2 \mathcal{S}_0\left(A_1+A_2\right)\left|q_L\right|.
\end{equation}
Indeed, by using this expression, it is immediate to show that for any drive applied to our system the following relation holds
\begin{equation}
\Delta \mathcal{S}_C=2 \mathcal{S}_0(A_1+A_2)\Delta N_{exc}.
\end{equation} 
\begin{figure}
	\centering
	\includegraphics[width=\linewidth]{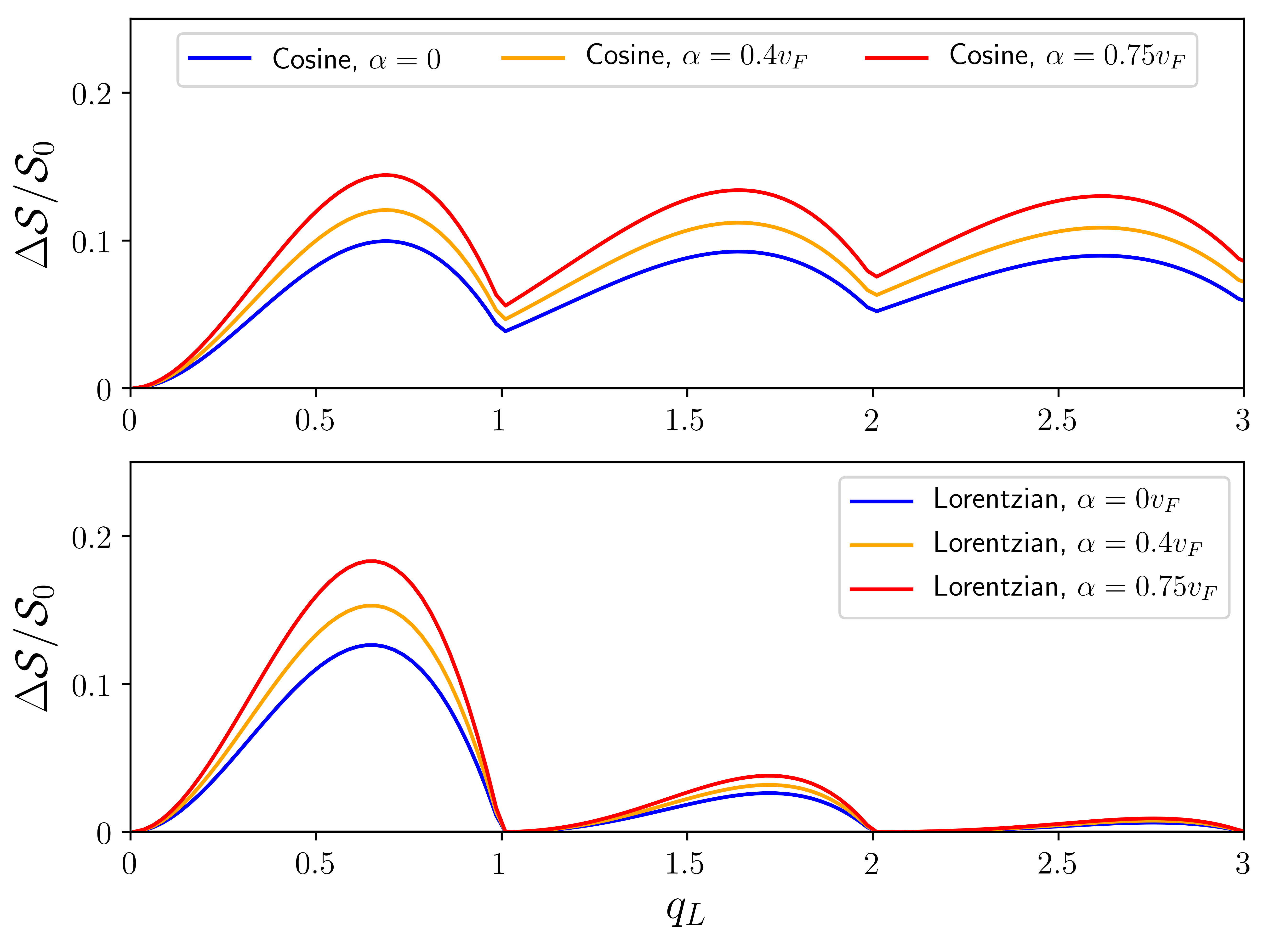}
	\caption{(Color online) Excess noise at zero temperature for a cosine drive (upper panel) and a periodic train of Lorentzian pulses (lower panel), for three values of Rashba spin-orbit strength, namely $\alpha=0 v_F$ (blue line), $\alpha=0.4 v_F$ (yellow line) and $\alpha=0.75 v_F$ (red line), as a function of number of electronic charge $q_L$ emitted in each period. The other parameter are $\eta=0.1$ and $\overline{\Delta}=0.75 v_F$.}
	\label{fig:Fig1}
\end{figure}The excess noise at zero temperature for a cosine drive and a periodic train of Lorentzian pulses is plotted in Fig.~\ref{fig:Fig1} for three values of Rashba spin-orbit strength. The width of each Lorentzian pulse is set to $\eta=0.1$, compatible with the range of parameters already accessible to experiments. Let us begin by commenting the excess noise for $\alpha=0$ (blue line in each panel), when the spin of electrons is perfectly oriented according to spin-momentum locking of helical edge states. Clearly, this curves for both cosine and Lorentzian drive have minima at integer values of $q_L$. The curve for the cosine drive always have a non-zero excess noise, showing that it is not an optimal drive since it inevitably generates additional and unwanted excitations. On the contrary, the excess noise for the Lorentzian drive drops to zero in correspondence with integer values of $q_L$, thus showing that Levitons are minimal excitations state for helical edge states. It is important to stress that in our proposed setup, the latter property could be tested by using a thin superconductor (instead of a so-far considered QPC geometry) employed as a beam-splitter in the HBT configuration. It is interesting to point out that a qualitatively similar behaviour of excess noise can be observed in a setup where the scattering mechanism is induced by a magnetic impurity coupling the two edge states, even though with less experimental control with respect to the presented superconducting configuration. \\ By moving on to the other curves in Fig.~\ref{fig:Fig1}, one can see that the values of the minima occurring for the cosine at integer values of $q_L$ depend on Rashba interaction strength and becomes bigger for increased values of $\alpha$. Interestingly, in the case of Lorentzian-shaped pulses, all these curves at finite values of $\alpha$ vanish when $q_L$ is an integer, thus showing that Levitons keep their nature of minimal excitation states even in the presence of Rashba spin-orbit coupling and the spin quantization axis is no more present. \\ 
\subsection{Hong-Ou-Mandel setup with Rashba interaction} 
\begin{figure}
\centering
\includegraphics[width=\linewidth]{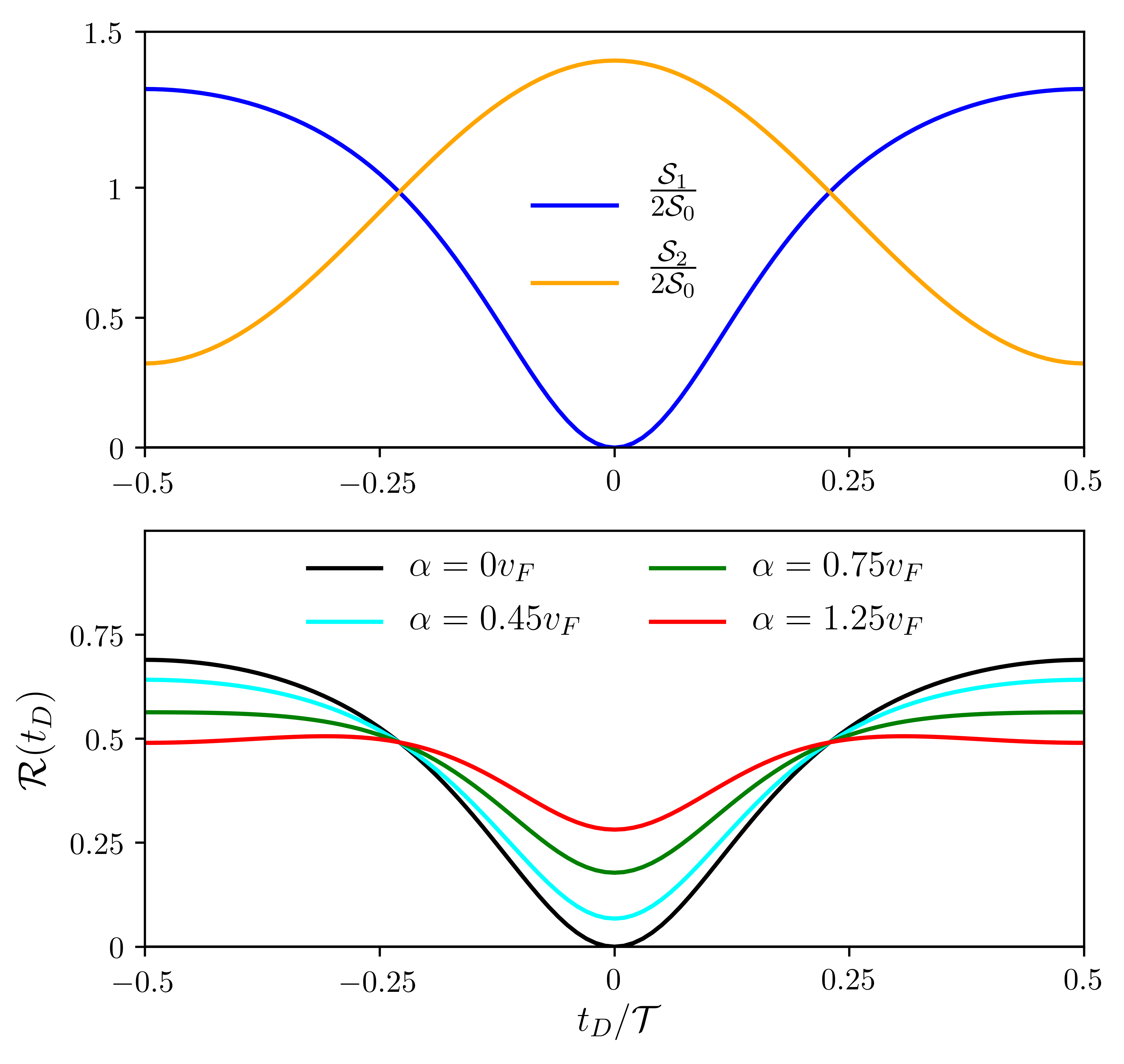}
\caption{(Color online) (Upper panel) Separate contributions $\mathcal{S}_1$ (blue lines) and $\mathcal{S}_2$ (yellow lines) as a function of $t_D$ for a Lorentzian drive at $q=1$. (Lower panel) HOM ratio $\mathcal{R}$ as a function of the time delay $t_D$ for $q=1$ with different values of $\alpha$. The other parameter are $\eta=0.1$, $\theta=0.01\omega$ and $\overline{\Delta}=0.75 v_F$.}
\label{fig:Fig4}
\end{figure}

Another interesting configuration is the Hong-Ou-Mandel (HOM) setup, which provides a way to characterize the properties of excitations emitted by a voltage source in the time domain. Hereafter, we investigate how to extend this configuration in the case of the thin superconducting terminal and discuss how the presence of Rashba interaction will affect the outcome.\\ In its usual electronic implementation, two identical voltage sources are applied to a system, with a finite delay $t_D$ between them. The emitted particles collide at a QPC and generate a noise with a specific behaviour as a function of the delay time. In particular, in the case of two identical colliding fermionic particles, the charge current fluctuations associated with the correlations of particle entering the reservoirs is exactly zero at $t_D=0$ due to completely destructive interference related to Pauli exclusion principle. Conversely, when two particle with opposite charge, i.e an electron and a hole, collide a maximum in the charge noise is present for null time delay.\\ The HOM interferometer can be reproduced in our setup, even without Rashba spin-orbit coupling. In this case, the only contribution to noise in addition to thermal noise is given by $\mathcal{S}_1$, which is generated by correlations between particles entering the reservoirs with opposite charges. Given this property, an electron-electron HOM interference pattern can be produced in our setup by focusing quantized Lorentzian-pulses and considering the configuration where $V_L(t)=V_{\rm lor}(t)$ and $V_{R}(t)=-V_{L}(t+t_D)$. In this case, one has that $q_L=-q_R\equiv q$, with $q>0$ an integer number, meaning that electrons and holes are emitted, respectively, by the left and the right contacts. Since $\mathcal{S}_1$ is sensitive to scattering processes where one of the two incoming particles reverse its charge, the correlations generated in this configuration are between particles entering the reservoirs with the same charge, giving rise to an electron-electron interference pattern when the delay $t_D$ is varied. Interestingly, deviations from this result can arise when $\alpha$ is finite and they can be considered as a signature for the presence of Rashba interaction. Indeed, $\mathcal{S}_2$ is associated with processes that always switch the sign of incoming particles, thus giving rise in our configuration to an electron-hole interference pattern.\\
In the upper panels of Fig.~\ref{fig:Fig4}, we plotted the two contributions to charge noise for a Lorentzian drive with $q=1$, as a function of the delay between the two voltages. Indeed, the convexity of the two contributions is opposite and, in particular, $\mathcal{S}_1$ is zero for $t_D=0$, while $\mathcal{S}_2$ displays a global maximum. This difference confirms that these two noise contributions correspond, respectively, to the interference between two identical electrons and between an electron and a hole. The simultaneous presence of these two types of interference is a peculiarity of our setup and cannot be observed in an analogous configuration where, for instance, tunnelling between edge states is induced by a magnetic impurity. As a result, different peculiar behaviour of the total charge noise can arise when Rashba spin-orbit strength is varied. For this reason, we inspect the outcome of this HOM experiment for different values of $\alpha$. 
Since the two contributions are not observable separately in a real experiment, one has to focus on the total charge noise. In order to make contact with previous literature on electronic HOM experiments, let us define the following ratio
\begin{equation}
\mathcal{R}(t_D)=\frac{\mathcal{S}^{HOM}_C-\mathcal{S}_C^{(0)}}{2\mathcal{S}_C^{HBT}-2\mathcal{S}_C^{(0)}},\label{eq:Ratio}
\end{equation}
where $\mathcal{S}^{HOM}_C=\mathcal{S}_C\left[V_L(t),-V_L(t+t_D)\right]$ is the noise in the HOM configuration and $\mathcal{S}_C^{(0)}$ is the noise in the absence of external drive given by
\begin{equation}
\mathcal{S}_C^{(0)}=\mathcal{S}_C\left[0,0\right]=2 (A_{\theta}+A_1+A_2) \mathcal{S}_0k_{\rm B} \theta.
\end{equation}This HOM ratio is plotted in the lower panel of Fig.~\ref{fig:Fig4} for a Lorentzian drive with $q=1$ with different values of $\alpha$ in a range of $t_D$ corresponding to a period $\mathcal{T}$. First of all, let us observe that at $\alpha=0$ (black line) HOM ratio vanishes at $t_D=0$. As long as Rashba interaction is present, the dip never vanishes and the value of HOM ratio at $t_D=0$ grows while increasing the value of $\alpha$. Moreover, the behaviour of $\mathcal{R}$ also varies significantly with Rashba strength. For finite values of Rashba strength, the ratio always presents a non-vanishing dip at $t_D=0$, whose value is monotonically increasing with $\alpha$. On the contrary, the value of the ratio is reduced at the ends of the period, where it becomes flatten as Rashba spin-orbit is stronger. Let us comment that a non-vanishing HOM dip would be present even in the configuration with $V_{R}(t)=V_{L}(t+t_D)$, where electrons are emitted from each source. In contrast with the other configuration, the non-vanishing dip is present even for a null Rashba interaction strength and is enhanced by increasing the value of $\alpha$. It is worth to underline that this is the first time a non-vanishing dip is predicted for in a symmetric HOM configuration with Levitons.\\ Interestingly, the existence of a non-vanishing dip in the HOM ratio appearing in Fig.~\ref{fig:Fig4} for finite values of $\alpha$ could be used as a way to assess the presence of Rashba spin-orbit coupling in helical edge states of a 2D TI. \\
\begin{figure}
	\centering
	\includegraphics[width=\linewidth]{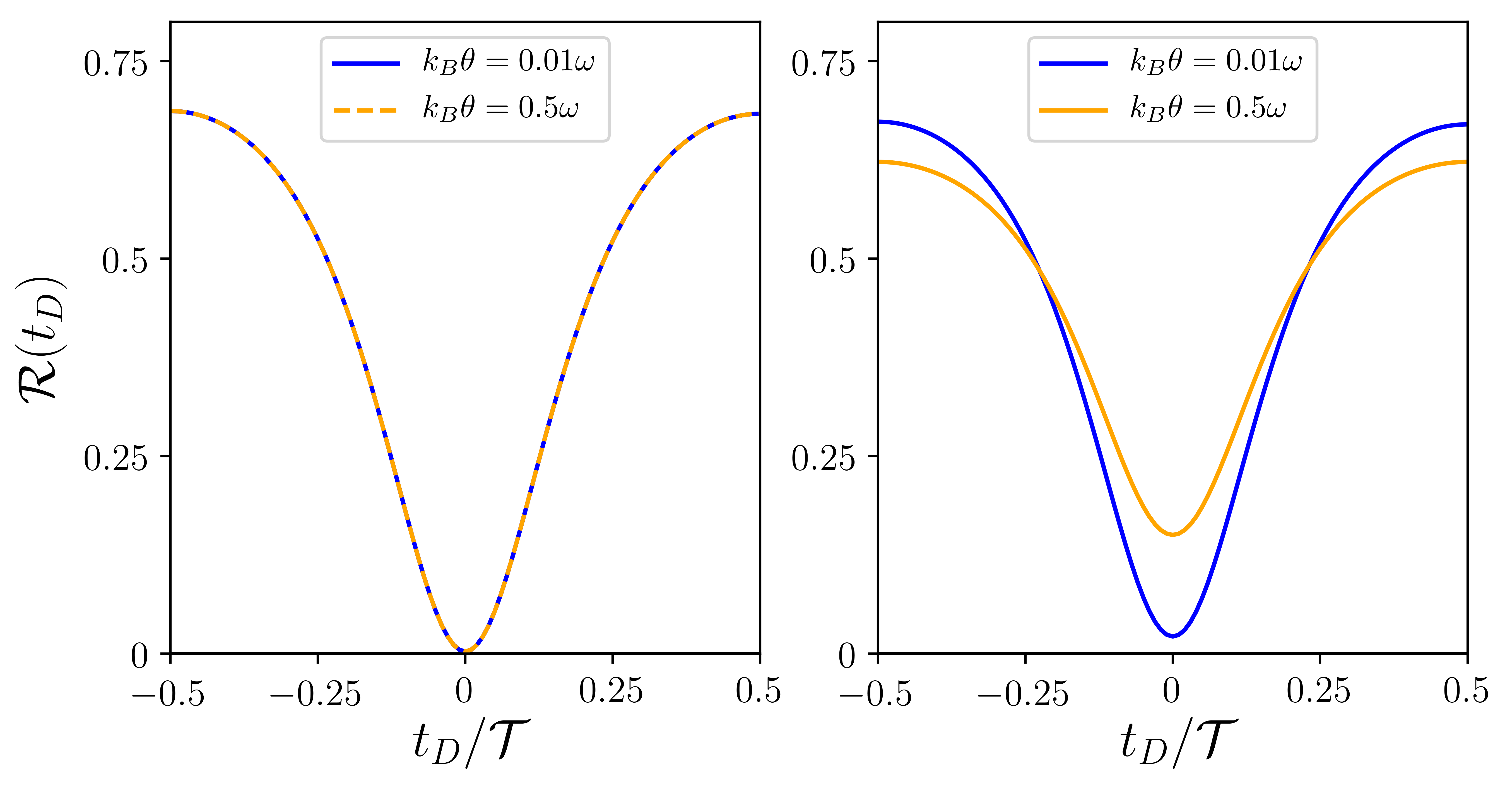}
	\caption{(Color online) HOM ratio $\mathcal{R}^{HOM}$ as a function of the time delay $t_D$ for $q=1$ and temperatures $\theta=0.01\omega$ (blue lines) and $\theta=0.5\omega$ (yellow lines). The case of $\alpha=0$ (left panel) and $\alpha=0.5 v_F$ (right panel) are compared. The other parameter are $\eta=0.1$ and $\overline{\Delta}=0.3 v_F$.}
	\label{fig:Fig3}
\end{figure}Finally, it is interesting to investigate the behaviour of HOM ratio as a function of temperature in the case of $q=1$. Previous papers proved that for this value of $q$ the ratio of HOM charge noise in a QPC geometry is independent of temperatures and acquire an universal analytical expression~\cite{dubois13,dubois13-nature,Rech16,Ronetti18}.\\ In order to perform a similar comparison for our system, we present in Fig.~\ref{fig:Fig3} the HOM ratio with two temperatures $\theta=0.01\omega$ (blue line) and $\theta=0.5\omega$ (yellow lines) in the absence (left panel) and presence (right panel) of Rashba spin-orbit coupling. In the first case, one can immediately see that HOM ratio is completely unaffected by temperature, even if the two chosen values for $\theta$ are very different. This universality is a consequence of the fact that, when $\alpha=0$, our setup with proximity induced superconductivity reproduces exactly the charge noise generated by two identical Levitons colliding at a conventional QPC. Nevertheless, the two curves in the right panel of Fig.~\ref{fig:Fig3} for $\alpha=0.5 v_F$ do not match anymore. The reason for the absence of universality is that $\mathcal{S}_2$ starts contributing when Rashba spin-orbit coupling is present. This term is due to the interference between an electron and a hole, which is not universal anymore with respect to temperature variations. Therefore, HOM ratio is the sum of two terms, one which is universal, namely $\mathcal{S}_1$ and the other one which is not, namely $\mathcal{S}_2$. Therefore, we can concluded that, when Rashba interaction are present, the universality of $\mathcal{R}$ with respect to temperature is not valid anymore. In passing, let us observe that, interestingly, the value of HOM dip is increased as the system temperature is risen.
\section{Conclusions \label{sec:conlusions}}
In this work, we have considered transport properties of Lorentzian-shaped voltage pulses in a single edge of a two-dimensional topological insulators, in presence of Rashba spin-orbit coupling. The two counterpropagating edge states can be connected with a thin superconductor, tunnel coupled at given position, which induces proximity correlations along them. Here, we have shown that this setup can be used to perform the analog of electron quantum optics experiments, where the role of a beamsplitter usually played by a quantum point contact is replaced by the thin superconductors. By computing charge noise, we have demonstrated that a HBT-like configuration with a single voltage could be used to demonstrate that Levitons are minimal excitations for charge transport also for helical edge states in presence of Rashba interaction, and it could be tested using scattering induced by superconducting proximity effect.
We have also considered a HOM interferometer, where the system is driven by two identical source of Levitons with opposite sign and delayed by a constant and tunable time. In this case, the presence of Rashba interaction could be verified by a non-vanishing HOM dip at zero delay. This result is a consequence of the fact that the charge noise is the sum of two terms, one corresponding to electron-electron interference and the other to electron-hole interference. Another interesting consequence of this property is that the HOM ratio for the case of the injection of a single Leviton is no longer universal with respect to temperature, in contrast with previous result for HOM experiments in two-dimensional electron gases or theoretical predictions for quantum Hall systems.
\begin{acknowledgments}
M. C. acknowledges support from the Quant-EraNet project Supertop.
\end{acknowledgments}
\appendix
\section{Detailed calculation of charge noise \label{app:noise}}
In this Appendix, we provide a detailed calculation of charge noise presented in Sec. \ref{sec:noise}. According to the relation in Eq.~\eqref{eq:noise_rel} We focus on the crosscorrelator $\mathcal{S}_{RL}$, which has been defined in the main text as $\mathcal{S}_{C}$. Its expression in terms of current operator $J_R$ (see Eq.~\eqref{eq:curr}) is
\begin{equation}
\mathcal{S}_{C}=\int_{0}^{\mathcal{T}}\frac{dt}{\mathcal{T}}\int dt'\left[\left\langle J_{R}(t')J_{L}(t)\right\rangle-\left\langle J_{R}(t')\right\rangle \left\langle J_{L}(t)\right\rangle\right],
\end{equation}
Before expressing charge noise in terms of fermionic operators in energy space, it is useful to change the notation defining outgoing and incoming fermionic operators as
\begin{align}
\mathbf{d}^T(\epsilon)&=\left(d_{+}(\epsilon), d_{-}(\epsilon), d^{\dagger}_{+}(-\epsilon), d^{\dagger}_{-}(-\epsilon)\right)\nonumber\\&=\left(d_1(\epsilon),d_2(\epsilon),d_3(\epsilon),d_4(\epsilon)\right),\\
\mathbf{c}^T(\epsilon)&=\left(c_{+}(\epsilon), c_{-}(\epsilon), c^{\dagger}_{+}(-\epsilon), c^{\dagger}_{-}(-\epsilon)\right)\nonumber\\&=\left(c_1(\epsilon),c_2(\epsilon),c_3(\epsilon),c_4(\epsilon)\right).
\end{align}
With this new notation, Eq.~\eqref{eq:scattering} can be recast as
\begin{equation}
d_i(\epsilon)=\sum_{j=1}^{4}\mathcal{M}_{ij}c_j(\epsilon),\hspace{5mm} i=1,\dots,4.\label{eq:di}
\end{equation}
By using Eq.~\eqref{eq:di}, charge noise can be expressed in a compact form in terms of scattering matrix and incoming operators in energy space as
\begin{widetext}
	\begin{align}
	\mathcal{S}_C&=\mathcal{S}_0\int \frac{dE'}{2\pi} \int \frac{dE}{2\pi}\sum_{i,j,k,l}\mathcal{M}_{3i}\mathcal{M}_{4k}\mathcal{M}_{2l}\mathcal{M}_{1j}\left(\left\langle c_{i}(-E')c_j(E')c_k(-E)c_l(E)\right\rangle-\left\langle c_{i}(-E')c_j(E')\right\rangle \left\langle c_k(-E)c_l(E)\right\rangle\right)+\nonumber\\&+2\sum_{i,j}\mathcal{M}_{4i}\mathcal{M}_{2j}\left(\left\langle c_{i}(-E')c_j(E')c_4(-E)c_2(E)\right\rangle-\left\langle c_{i}(-E')c_j(E')\right\rangle \left\langle c_4(-E)c_2(E)\right\rangle\right)+\nonumber\\&+2\sum_{i,j}\mathcal{M}_{3i}\mathcal{M}_{1j}\left(\left\langle c_{i}(-E')c_j(E')c_3(-E)c_1(E)\right\rangle-\left\langle c_{i}(-E')c_j(E')\right\rangle \left\langle c_3(-E)c_1(E)\right\rangle\right),
	\end{align}
\end{widetext}
where we defined $\mathcal{S}_0=\frac{e^2}{\mathcal{T}}$.
Since, the four-operator averages can be simplified into two-operator averages by resorting to Wick-theorem, we define a $4\times 4$ matrix which contains all average values of operators $c_{\pm}(E)$ (see Eqs. \eqref{eq:av_c1} and \eqref{eq:av_c2}) 
\begin{equation}
\mathcal{F}_{ij}(E',E)=\langle c_{i}(E')c_j(E)\rangle,
\end{equation}
whose elements are finite only if $\left|i-j\right|=2$. By using Wick's theorem and matrix $\mathcal{F}$ the noise becomes
\begin{widetext}
	\begin{align}
	\mathcal{S}_C&=\mathcal{S}_0\int \frac{dE'}{2\pi} \int \frac{dE}{2\pi}\Bigg[\sum_{i,j,k,l}\mathcal{M}_{3i}\mathcal{M}_{4k}\mathcal{M}_{2l}\mathcal{M}_{1j}\left(\mathcal{F}_{il}(-E',E)\mathcal{F}_{jk}(E',-E)-\mathcal{F}_{ik}(-E',-E)\mathcal{F}_{jl}(E',E)\right)+\nonumber\\&-2\sum_{i,j}\mathcal{M}_{4i}\mathcal{M}_{2j}\left(\mathcal{F}_{i4}(-E',-E)\mathcal{F}_{j2}(E',E)-\mathcal{F}_{i2}(-E',E)\mathcal{F}_{j4}(E',-E)\right)+\nonumber\\&-2\sum_{i,j}\mathcal{M}_{3i}\mathcal{M}_{1j}\left(\mathcal{F}_{i3}(-E',-E)\mathcal{F}_{j1}(E',E)-\mathcal{F}_{i1}(-E',E)\mathcal{F}_{j3}(E',-E)\right)\Bigg].
	\end{align}
\end{widetext}
By imposing the constraint $\left|i-j\right|=2$ valid for the element of matrix $\mathcal{F}$ and restoring the original notation for fermionic operators, the above expression becomes
\begin{widetext}
	\begin{align}
	\mathcal{S}_C&=\mathcal{S}_0\int \frac{dE'}{2\pi} \int \frac{dE}{2\pi}\nonumber\times\\&\Bigg[2\mathcal{M}_{41}\mathcal{M}_{33}\mathcal{M}_{21}\mathcal{M}_{13}\langle c_{+}(E')c^{\dagger}_+(E)\rangle\langle c_{+}(-E')c^{\dagger}_+(-E)\rangle+2\mathcal{M}_{33}\mathcal{M}_{41}\mathcal{M}_{13}\mathcal{M}_{21}\langle c^{\dagger}_{+}(E')c_+(E)\rangle\langle c^{\dagger}_{+}(-E')c_+(-E)\rangle+\nonumber\\&+\left(\mathcal{M}_{33}\mathcal{M}_{41}\mathcal{M}_{23}\mathcal{M}_{11}+\mathcal{M}_{33}\mathcal{M}_{43}\mathcal{M}_{21}\mathcal{M}_{11}+\mathcal{M}_{31}\mathcal{M}_{43}\mathcal{M}_{13}\mathcal{M}_{21}+\mathcal{M}_{31}\mathcal{M}_{41}\mathcal{M}_{13}\mathcal{M}_{23}\right)\langle c_{+}(E')c^{\dagger}_+(E)\rangle\langle c^{\dagger}_{+}(E')c_+(E)\rangle+\nonumber\\&+2\mathcal{M}_{42}\mathcal{M}_{34}\mathcal{M}_{22}\mathcal{M}_{14}\langle c_{-}(E')c^{\dagger}_-(E)\rangle\langle c_{-}(-E')c^{\dagger}_-(-E)\rangle+2\mathcal{M}_{34}\mathcal{M}_{42}\mathcal{M}_{14}\mathcal{M}_{22}\langle c^{\dagger}_{-}(E')c_-(E)\rangle\langle c^{\dagger}_{-}(-E')c_-(-E)\rangle+\nonumber\\&+\left(\mathcal{M}_{34}\mathcal{M}_{42}\mathcal{M}_{24}\mathcal{M}_{12}+\mathcal{M}_{34}\mathcal{M}_{44}\mathcal{M}_{22}\mathcal{M}_{12}+\mathcal{M}_{32}\mathcal{M}_{44}\mathcal{M}_{14}\mathcal{M}_{22}+\mathcal{M}_{32}\mathcal{M}_{42}\mathcal{M}_{14}\mathcal{M}_{24}\right)\langle c_{-}(E')c^{\dagger}_-(E)\rangle\langle c^{\dagger}_{-}(E')c_-(E)\rangle+\nonumber\\&+\left(\mathcal{M}_{31}\mathcal{M}_{43}\mathcal{M}_{14}\mathcal{M}_{22}+\mathcal{M}_{31}\mathcal{M}_{42}\mathcal{M}_{14}\mathcal{M}_{23}+\mathcal{M}_{34}\mathcal{M}_{42}\mathcal{M}_{11}\mathcal{M}_{23}+\mathcal{M}_{34}\mathcal{M}_{43}\mathcal{M}_{11}\mathcal{M}_{22}\right)\langle c_{+}(E')c^{\dagger}_+(E)\rangle\langle c^{\dagger}_{-}(E')c_-(E)\rangle\nonumber+\\&+\left(\mathcal{M}_{33}\mathcal{M}_{41}\mathcal{M}_{24}\mathcal{M}_{12}+\mathcal{M}_{33}\mathcal{M}_{44}\mathcal{M}_{12}\mathcal{M}_{21}+\mathcal{M}_{32}\mathcal{M}_{44}\mathcal{M}_{13}\mathcal{M}_{21}+\mathcal{M}_{32}\mathcal{M}_{41}\mathcal{M}_{13}\mathcal{M}_{24}\right)\langle c_{-}(E')c^{\dagger}_-(E)\rangle\langle c^{\dagger}_{+}(E')c_+(E)\rangle+\nonumber\\&+\left(\mathcal{M}_{31}\mathcal{M}_{43}\mathcal{M}_{12}\mathcal{M}_{24}+\mathcal{M}_{31}\mathcal{M}_{44}\mathcal{M}_{12}\mathcal{M}_{23}+\mathcal{M}_{32}\mathcal{M}_{44}\mathcal{M}_{11}\mathcal{M}_{23}+\mathcal{M}_{32}\mathcal{M}_{43}\mathcal{M}_{11}\mathcal{M}_{24}\right)\langle c_{+}(E')c^{\dagger}_+(E)\rangle\langle c_{-}(-E')c^{\dagger}_-(-E)\rangle+\nonumber\\&+\left(\mathcal{M}_{33}\mathcal{M}_{41}\mathcal{M}_{22}\mathcal{M}_{14}+\mathcal{M}_{33}\mathcal{M}_{42}\mathcal{M}_{14}\mathcal{M}_{21}+\mathcal{M}_{34}\mathcal{M}_{41}\mathcal{M}_{13}\mathcal{M}_{22}+\mathcal{M}_{34}\mathcal{M}_{42}\mathcal{M}_{13}\mathcal{M}_{21}\right)\langle c^{\dagger}_{-}(-E')c_-(-E)\rangle\langle c^{\dagger}_{+}(E')c_+(E)\rangle+\nonumber\\&-2\left(\mathcal{M}_{42}\mathcal{M}_{24}+\mathcal{M}_{44}\mathcal{M}_{22}\right)\langle c_{-}(E')c^{\dagger}_-(E)\rangle\langle c^{\dagger}_{-}(E')c_-(E)\rangle+\nonumber\\&-2\left(\mathcal{M}_{31}\mathcal{M}_{13}+\mathcal{M}_{33}\mathcal{M}_{11}\right)\langle c_{+}(E')c^{\dagger}_+(E)\rangle\langle c^{\dagger}_{+}(E')c_+(E)\rangle\Bigg].
	\end{align}
\end{widetext}
Finally, by using the expression for scattering matrix in Eq.~\eqref{eq:scattering_m1}, one has
\begin{align}
&\mathcal{M}_{11}=\mathcal{M}_{22}=\mathcal{M}_{33}=\mathcal{M}_{44}=t_{ee},\\&\mathcal{M}_{13}=\mathcal{M}_{24}=\mathcal{M}_{31}=\mathcal{M}_{42}=t_{eh},\\
&\mathcal{M}_{12}=\mathcal{M}_{21}=\mathcal{M}_{34}=\mathcal{M}_{43}=0,\\
&\mathcal{M}_{14}=\mathcal{M}_{23}=\mathcal{M}_{32}^{*}=\mathcal{M}_{41}^{*}=r_{eh},
\end{align}
one finds again the expression reported in the main text.

\end{document}